# Coexistance of non-Fermi liquid behavior and bi-quadratic exchange coupling in La-substituted CeGe: Non-linear susceptibility and DFT + DMFT study

Karan Singh[1], Antik Sihi[1], Sudhir K. Pandey[2] and K. Mukherjee[1]

[1]School of Basic Sciences, Indian Institute of Technology Mandi, Mandi 175075, Himachal Pradesh, India

[2]School of Engineering, Indian Institute of Technology Mandi, Mandi 175075, Himachal Pradesh, India

## Abstract

Studies connected with the investigations of "non-Fermi liquid" (NFL) systems continue to attract interest in condensed matter physics community. Understanding the anomalous physical properties exhibited by such systems and its related electronic structures is one of the central research topics in this area. In this context, Ce-based and Ce-site diluted (with non-magnetic ions) compounds provide a fertile playground. Here, we present a detailed study of non-linear DC susceptibility and combined density functional theory plus dynamical mean field theory (DFT+DMFT) on $Ce_{0.24}La_{0.76}Ge$. Theoretical investigation of $4f$ partial density of states, local susceptibility and self-energy demonstrates the presence of NFL behavior which is associated with fluctuating local moments. Non-linear DC susceptibility studies on this compound reveal that the transition from NFL state to the new phase is due to development of the bi-quadratic exchange coupling and it obeys the non-linear susceptibility scaling. Under the application of magnetic fields, local moments interact spatially through conduction electrons resulting in magnetic fluctuations. Our studies point to the fact that the origin of the observed bi-quadratic exchange coupling is due to the spatial magnetic fluctuations.

## 1. Introduction

Novel electronic states of strongly correlated electron systems with partially filled $f$ or $d$-shells continue to invoke interest in the condensed matter community [1]. In many members of this family, non-Fermi liquid (NFL) phenomenon is often developed in the vicinity of a quantum critical point (QCP). At QCP, magnetic ordering temperature is driven to zero by alloying or by applying pressure or magnetic fields [2, 3]. Some of the best-known examples in the correlated $f$ and $d$ electrons systems where the NFL behavior is observed are $CeCoIn_5$ [4], $CeCu_{5.1}Au_{0.9}$ [5], $Ce(Ru_{0.5}Rh_{0.5})_2Si_2$ [6], $YbRh_2Si_2$ [7], $UCu_{5-x}Pd_x$ [8], and $Ni_xPd_{1-x}$ [9] etc. The low temperature physical properties of correlated electron systems usually follow the Landau's Fermi-liquid (LFL) theory which is based on a paradigm that the $T^2$ dependence of DC resistivity ($\rho$) is due to prevailing single transport relaxation rate, which, originates from the electron-electron scattering. When the Fermi liquid state becomes unstable, an anomalous state (with $\rho$ varying as $T^\alpha$, with $\alpha \neq 2$), i.e. the so-called NFL behavior would emerge. Such a behavior arises due to the presence of quantum fluctuations, which, are enhanced due to a strong competition between on-site Kondo effect and the inter-site Ruderman-Kittel-Kasuya-Yosida (RKKY) interactions. Additionally, some attempts have also been made to explain the NFL behavior based on the possibility of an unconventional Kondo effect [10] or disorder effects [11]. Generally, clean strongly correlated systems are characterized by large amount of the scattering of the metallic carries by the localized magnetic moments. When the temperature is lowered, the lattice translation invariance set in and a quick drop of the resistivity occurs below so-called coherence temperature [12]. The dilution of localized magnetic moments by substitution of non-magnetic ion has led to a single-impurity like behavior and coherent temperature is affected [13]. Furthermore, it is expected that the interplay of the disorder effect would destroy the lattice translation invariance. Thereby the coherent temperature is suppressed and might lead to the observation of NFL behavior [12, 14]. As a result, various anomalous variations of the physical properties are observed due to the presence of different relaxation rates. Hence, it can be said that in presence of disorder, the interplay of conduction electrons with localized magnetic moments (through Kondo coupling) is affected, which might give rise to the NFL behavior [14]. This mechanism can be understood on the basis that disorder lead to a broad distribution of Kondo temperature. It results in a finite fraction of the unquenched localized moments which resides at all temperatures. This adds an exchange interaction term in the Hamiltonian, which can be due to the randomness in the Ruderman-Kittel-Kasuya-Yosida (RKKY) intersite coupling [3]. This phenomenon is usually expected in disordered Kondo alloy, where the intersite interaction energy dominates over the Kondo energy associated with magnetic fluctuations [15]. These fluctuations may develop bi-quadratic exchange coupling at very low temperatures and/or under the application of external magnetic fields [16, 17, and 18].

In some $d$-electron systems (like $BaRuO_3$, $SrRuO_3$) exhibiting NFL behavior, the electronic properties has been systematically studied by employing a single-site dynamical mean-field model with rotationally invariant interactions. For such model, Matsubara self-energy (Im$\sum(i\omega)$) varies as $\omega^\alpha$ (where $\omega$ = frequency and exponent $\alpha = \frac{1}{2}$) [19, 20]. R. H. McKenzie *et al*., had reported that the presence of NFL state gives rise to a positive slope in the real part of the self-energy and exhibits a dip in the imaginary part of the self-energy [21]. In the presence of magnetic fields, Fermi liquid behavior can be observed with $\alpha = 1$ [19, 20]. In $4f$-electron systems, density functional theory plus dynamical mean field theory (DFT+DMFT) approach has been applied to understand the electronic structure [22, 23]. But, to the best of our knowledge, a systematic detailed investigation of electronic structure along with its relationship with the bi-quadratic exchange coupling is lacking for the $4f$-based NFL systems.

To shed some light on the possible presence of bi-quadratic exchange coupling, associated with NFL behavior, we choose our recently studied $Ce_{0.24}La_{0.76}Ge$ compound [24]. The corresponding parent compound, CeGe undergo antiferromagnetic type transition along with development of multipolar moments around 10.7 K. In the presence of magnetic fields, long-ranged antiferromagnetic order is preserved. The Sommerfeld coefficient of this compound is found to be around ~ 0.267 J/mol-$K^2$ [25]. This suggests an increase in the mass of electrons by partial screening of $4f$ moments through the conduction electrons. Hence, this compound can be considered in the category of disordered Kondo system. The above statement can be made on the basis of the fact that the Kondo peak in resistivity is absent [25-28]. This observation is ascribed to the dominating intersite interactions energy, which impedes the whole screening of magnetic moments. Due to the substitution of La, disorder driven NFL behavior is observed in $Ce_{0.24}La_{0.76}Ge$ [24]. Interestingly, in this compound, in the presence of magnetic fields, an anomaly (above 2 ($10^4$) Oe) in heat capacity, is noted over a wide temperature range. The peak got shifted to higher temperatures with increasing fields [24]. Thus, in this compound a changeover from NFL state to a new phase is noted. These observations raises the following questions: i) whether the NFL state is associated with the local moment fluctuations? ii) whether the observed new phase is due to presence of bi-quadratic exchange coupling? iii) If yes, is the bi-quadratic exchange coupling is related to fluctuations?

In order to address these queries, in this manuscript, we present a detailed study of non-linear DC susceptibility and combined density functional theory plus dynamical mean field theory (DFT+DMFT) on $Ce_{0.24}La_{0.76}Ge$. Non-linear DC susceptibility provides an ideal method to test the equilibrium nature of a phase transition involving higher order exchange interaction terms. The scaling of non-linear susceptibility is a signature of the genuine existence of equilibrium phase transition where static critical exponents are found and they are related to each other through hyperscaling relation [29-31]. A reliable explanation about the NFL behavior can be given, with the aid of DFT+DMFT calculations [32]. This approach solves the localized impurity problem at finite temperature, in which, the self-energy is the function of $\omega$ only. It also describes the localization effect and quasiparticle excitations [32, 33]. Additionally, to support computational methods, parameters adopted and the obtained results for this compound, we will also discuss the computational result of the parent CeGe.

Our results indicate that in $Ce_{0.24}La_{0.76}Ge$, non-linear DC susceptibility scaling is satisfied. The scaling suggests that transition from the NFL to a new phase is due to development of the bi-quadratic exchange coupling in the presence of magnetic fields. In zero magnetic field, theoretical results also indicate to the presence of NFL state which is associated with local moment fluctuations. These observations can be attributed to the fact that the local moments interact spatially through conduction electrons in the presence of magnetic fields. It results in the development of spatial magnetic fluctuations which is associated with the bi-quadratic exchange coupling. Thus, it can be said that this field induced phase is derived from the spatial magnetic fluctuations.

## 2. Methods

The compounds, CeGe and $Ce_{0.24}La_{0.76}Ge$ are the same as those reported in Refs [24, 25]. Non-linear susceptibility measurements are performed using Magnetic Property Measurement System (MPMS), from Quantum design, USA. For computational studies, we used dynamical mean field theory on top of density functional theory (DFT+DMFT). In recent years, the DFT+DMFT is one of the most advanced electronic structure method for investigations of strongly correlated electron systems. In this study, the spin-orbit coupled calculations are performed for these compounds with the help of full-potential linearized-augmented plane-wave method as implemented in WIEN2k code [34]. The local density approximation is chosen as the exchange-correlation functional for these calculations [35]. The DFT+DMFT calculations are done using eDMFT [32] code, where the DFT part is carried out using WIEN2k code [34]. In this scenario, the Hubbard U = 6.0 eV and Hund's coupling J = 0.7 eV are fixed to solve the impurity problem using continuous-time quantum Monte Carlo (CTQMC) method [36]. Here, 'nominal' method is selected to take care of the double-counting problem [37]. To transfer the spectral function from imaginary to real axis, maximum entropy analytical continuation method is employed [38]. 160 k-points in irreducible Brillouin zone is the set for this study. It is to be noted that both these compounds crystallize in the orthorhombic structure with *Pnma* space group. The lattice parameters used for the calculation are extracted from the Rietveld refinement of the X-ray diffraction pattern. At this point, it is to be mentioned that virtual crystal approximations [39] is chosen to study the effect of La-doped on the correlated Ce $4f$ electrons.

## 3. Results

### 3.1 Non-linear susceptibility study

The non-linear susceptibility is expressed as [18, 40]

$$M/H = \chi_0 - \chi_1 H^{a(T)-1} \quad \dots (1)$$

where $\chi_0$ is the linear magnetic susceptibility and $\chi_1$ is a non-linear magnetic susceptibility term. $a(T)$ is the exponent determined at different temperatures. In $Ce_{0.24}La_{0.76}Ge$, in order to investigate the possibility of the presence of higher order magnetization, the non-linear DC susceptibility is studied. This study is useful for the determination of high-order susceptibility terms which are dependent on very low frequencies (less than $10^{-2}$ Hz) and are generally hidden in direct AC susceptibility measurements. The non-linear parts of DC susceptibilities are obtained from field and temperature dependence of magnetization (see figure S1 of supplementary information) using the protocol as given in the Refs [18, 40, and 41]. Figure 1 shows the temperature ($T$) dependent $\chi_1$. Interestingly, it is noted that $\chi_1$ grows sharply below $T^*$ (~ 3 K). This observed feature can be referred to a new energy scale (shaded region in figure 1) due to the development of higher order susceptibility in the presence of magnetic fields [18]. Inset of figure 1 shows the temperature dependent exponent $a(T)$. This exponent decreases with decreasing temperature and reaches 1.5 around 3 K. This observed value is not in accordance to the quadrapole picture ($a(T) = 4$), octapole picture ($a(T) = 6$), etc. As mentioned earlier, in the zero field limit ($H \to 0$), disorders driven NFL behavior is present in this compound [24]. The variation of linear susceptibility $\chi_0$ (see figure S2 of supplementary information), indicates that there is the presence of local magnetic moments which increases with decreasing temperature. In the external magnetic fields, these moments are responsible for the development of a new phase which is possibly associated with bi-quadratic exchange coupling. This feature can be clearly seen as an anomaly in heat capacity measurements over a wide temperature range (around $T^*$), above 2 ($10^4$) Oe (see figure S3 of supplementary information). Inset of figure 1 (b) shows the total non-linear susceptibility $\chi_{nl}$ ($\chi_0 - M/H$) as a function of the magnetic field. In temperature range 4 – 10 K, it is noted that $\chi_{nl}$ is independent at low fields. Above 1 ($10^4$) Oe, it increases non-linearity and slope of the curve also increases with decreasing temperature. Hence, it might be said that a field induced bi-quadratic exchange coupling possibly leads to the development of the higher order magnetization.

To investigate the equilibrium nature of phase transition associated with bi-quadratic exchange coupling, we relate this coupling to the spin freezing which leads to the divergence of the so-called spin glass susceptibility $\chi_{SG}$ [42]

$$\chi_{nl} = \beta^3 \ (\chi_{SG} - 2/3) \ \dots \ (2)$$

where $\beta$ is critical exponent associated with an order parameter q. This order parameter q is proportional to $t^\beta$ ($t = (T/T^* - 1)$ is the reduced temperature). $\chi_{SG}$ is defined as [42]

$$\chi_{SG} = \beta^2 [(\langle S_i S_j \rangle - \langle S_i \rangle \langle S_j \rangle)^2]_{av} \ \dots \ (3)$$

where $\langle \dots \rangle$ and $[\dots]$ denotes to the thermal averages and an average over the disorders, respectively; $S_i$ and $S_j$ are the magnetic moment at site $i$ and $j$, respectively. Further, the most relevant test for ascribing the critical behavior of the phase transition can be extracted by the scaling of $\chi_{nl}$. In our case, it is anticipated that $\chi_{nl}$ ($T^*$, $H$) begins to diverge around temperature $T^*$ or below (where $\chi_0$ is non-divergent) with the power law as $\chi_{nl}$ ($T^*$, $H$) $\propto H^{2/\delta}$ and $t^{-\gamma}$ ($\delta$ and $\gamma$ are critical exponents that have related with $\beta$). A scaling form of $\chi_{nl}$ is defined as [40, 42].

$$\chi_{nl} \ (T^*, H) = H^{2/\delta} \ f \ (t/H^{2/\varphi}) \ \dots \ (4)$$

where $\varphi = \beta + \gamma$, $\delta = (\beta + \gamma)/\beta$ and f (x) is an arbitrary scaling function and follows the asymptotic behavior; f (x) = const, $x \to 0$ and f (x) = $x^{-\alpha}$, $x \to \infty$. The exponents $\delta$ and $\varphi$ can be determined through scaling. Figure 1 (b) shows the scaling of $\chi_{nl}$($T^*$, $H$) with parameters $T^*$ = 3 K, $\delta$ = 5±0.3 and $\varphi$ = 4±0.1, above 2 ($10^4$) Oe. Using relation $\varphi = \beta + \gamma$ and $\delta = (\beta + \gamma)/\beta$, the $\gamma$ is equals to $\varphi(\delta-1)/\delta$ and has a value of 3.2. A similar type of scaling and critical exponents has been reported for the GdAl system [18]. The critical exponent β (= φ/δ) associated with order parameter q is equals to 0.8 and determines the exponent ν of the correlation length ξ ($\propto$ t$^{-\nu}$) from the relation $d\nu = 2\beta + \gamma$ ($d$ = dimensionality) [43]. For $d$ = 3, ν equals to 1.6, and this value suggests that the order parameter can develop a spin-freezing like transition when bi-quadratic exchange coupling associated with squared correlation function $(S_i . S_j)^2$ turns into long-range ordering [42]. Hence, it can be said that the field induced new phase originate due to establishment of partial order parameter associated with bi-quadratic exchange coupling (further discussed in the section 4). In the next section, we perform DFT + DMFT calculations for this compound to get a deeper insight about the NFL state.

## 3.2 Computational study using DFT+DMFT

### 3.2.1 Density of states and susceptibility

DFT in combination with the DMFT can be used to capture the evolution of NFL state arising from 4$f$-electron configurations under finite temperatures [21, 22, and 32]. In this sub-section, we will focus on the computational results for CeGe and Ce$_{0.24}$Ce$_{0.76}$Ge. For CeGe, figure 2 (a) and (b) show the partial density of states (PDOS) for $J = 5/2$ and $J = 7/2$ sub-bands of Ce-4$f$ in energy window -10 to 10 eV, respectively. Ce-4$f$ orbitals are splited into the $J = 5/2$ and $J = 7/2$ sub-bands, which is induced by the spin-orbit coupling. In both cases, a large peak is observed at 2.90 eV (for $J = 5/2$) and 3.17 eV (for $J = 7/2$), which can be assigned to the upper Hubbard bands [23, 32]. These observed peaks can arise due to the localized 4$f$-electrons. The distance between two peaks is about 270 meV, which is close to the reported value of spin-orbit splitting $\Delta_{SO}$ = 280 meV [44]. A weak lower Hubbard band is also observed at the -1.7 eV (figure 2 (a)) and -0.5 eV (upper inset of figure 2 (b)) for $J = 5/2$ and $J = 7/2$, respectively. This observed upper and lower Hubbard band is arised due to the effective on-site Coulomb potential (U$_{eff}$). At low temperatures (around 30 K), similar upper Hubbard band with a shift in energy is seen, for both $J = 5/2$ and $J = 7/2$, but, the intensity of the peak is smaller. The lower Hubbard band disappears for $J = 5/2$, but, interestingly, a quasiparticle peak appears in the vicinity of Fermi level (see upper inset of figure 2 (a)). This means that U$_{eff}$ is reduced due to partially screening of the 4$f$-electrons. With decreasing temperatures, magnetic moments combine with conduction bands to form a heavy quasiparticle fluid, having a mass which is very large compared to the mass of free electrons. This forms a quasiparticle peak in the vicinity of Fermi level in Ce-4*f* PDOS. On the other hand, it reduces the weight of the upper Hubbard band and the weaker lower Hubbard band is suppressed. This indicates that $J = 5/2$ sub-band could be the source of the effective mass. A similar feature was also noted in other Ce-based compound like CeRu$_2$Si$_2$ [23]. Mostly, quasiparticles form the Fermi liquid state which has a mean free path much larger than their wavelength [45]. For this state, resistivity varies quadratically at low temperature and self-energy $\sum(\omega)$ varies as $\omega^2$ [46]. For CeGe, quadratic variation of resistivity has not been observed [25]; suggesting that the mean free path might be smaller than its wavelength. The quasiparticle picture is not seen for the $J = 7/2$ sub-band, but we see the three peaks in the energy window -0.2 to -0.35 eV (lower inset of figure 2 (b)). In this energy window, the $J = 5/2$ and $J = 7/2$ sub-bands can be further splited into doublet ($\Gamma_7$) and quartet state ($\Gamma_8$) under crystalline electric field effects [47, 48, and 49]. The $J = 5/2$ splits into ground state $\Gamma_8$ and excited state $\Gamma_7$ where $J = 7/2$ splits into ground state $\Gamma_7$ and two excited states $\Gamma_8$ and $\Gamma_7$. Lower inset of figure 2 (a) shows a peak in the energy window -0.2 to -0.35 eV, indicating that there might be absence of quartet state for 5/2 sub-bands. But, for $J = 7/2$, three peaks is observed in this energy window, indicating the presence of the quartet state. The distance between the ground state $\Gamma_7$ and first excited state $\Gamma_8$ and the distance between the first excited state $\Gamma_8$ and second excited state $\Gamma_7$ are 55 meV and 56 meV, respectively. The observed energy difference between $\Gamma_7$ and $\Gamma_8$ is close to the 50 meV which has been reported in other Ce-based compound; e.g.CeB$_6$ [48, 50]. The quartet state usually produces multipolar moments that have also been reported experimentally for CeGe [25].

Figure 2 (c) and 2 (d) shows the plot of the PDOS in the energy window -10 to 10 eV at 30 and 300 K for the $J = 5/2$ and $J = 7/2$ sub-bands, respectively, of Ce$_{0.24}$La$_{0.76}$Ge compound. It is noted that both sub-bands behavior is similar. But one large peak at 2.39 eV (away from the Fermi level) is observed. With decreasing temperature (30 K), this peak becomes some smaller and broader. Hence, it can be said that the localized nature 4*f* electrons (as appeared in CeGe) preserves in the Ce$_{0.24}$La$_{0.76}$Ge. The quasiparticle picture is not observed at the Fermi level in the Ce$_{0.24}$La$_{0.76}$Ge compound (Inset of figure 2 (c) and 2 (d)). It can be due to the presence of NFL state [24]. Hence, it can be said that dynamical screening effect is not important in this compound which possibly evades the quasiparticle picture. In Refs [51, 52] it has been discussed that the quasiparticle is suppressed in the NFL paradigm and some local moments are preserved.

Within the framework of DFT+DMFT we can compute the local spin susceptibility $\chi_{loc}$ for both compounds [53, 54, and 55]

$$\chi_{loc} = \int_0^\beta d\tau \langle S_i(\tau)S_i(0)\rangle \ \dots \ (5)$$

where $\beta$ is the inverse temperature. The operator $S_i = (1/2M)\sum_{\gamma=1}^{M}(c_{i\gamma\uparrow}^{\dagger} c_{i\gamma\uparrow} - c_{i\gamma\downarrow}^{\dagger} c_{i\gamma\downarrow})$, $M$ indicates to the number of $f$ orbitals, $i$ is the site index, and $\gamma$ is the orbital index. The operator $c_{i\gamma\uparrow}$ ($c_{i\gamma\uparrow}^{\dagger}$) is the annihilator (creator) of the electron. $\langle S_i(\tau)S_i(0)\rangle$ is the imaginary time dependent spin-spin correlation function. At high temperatures, $\chi_{loc}$ follow the behavior as; $\chi_{loc} = C/T$ (where $C$ = Curie constant) [55]. Figure 3 shows the temperature dependent $\chi_{loc}^{-1}$ for CeGe. The fitted $\chi_{loc}$ gives $C$ ($\sim\langle S_i(\tau)S_i(0)\rangle$) equal to 0.92. This implies that the $4f$ electron sustains the local nature of the magnetic moments and the spins at different $\tau$ points are correlated. The effective local magnetic moment $\mu_{eff}$ is also calculated using relation; $2.827\sqrt{C}$ $\mu_B$, which is around $2.71\mu_B$. This value is close to the expected value of $2.52\mu_B$, calculated using the formula $\mu_{eff} = g_J\sqrt{J(J+1)}$ $\mu_B$ where $J = 5/2$, $g_J = 6/7$, and reported experimental value ($\sim 2.4\mu_B$) for this compound [25]. For $Ce_{0.24}La_{0.76}Ge$, the computed $\chi_{loc}$ does not vary significantly with temperature. This might be due to trivial presence of the spin–spin correlations $\langle S_i(\tau)S_i(0)$. This suggests that local moment fluctuations are large implying that spins at different $\tau$ points do not show temperature dependent correlations. As observed experimentally, temperature dependence data indicate the presence of the spin-spin correlations in this compound which is significantly weaker in comparison to CeGe. This suggests that virtual crystal approximation is unable to capture spin-spin correlations exactly. Moreover, experimental data also suggest the non-local spin-spin interactions. Even though single site DMFT gives the direct information about the local interactions, however, one expects that hybridization function may contain indirect information about the non-local spin-spin interactions when calculations will be performed on the structure with two or more than two inequivalent impurity sites containing Ce atoms. In order to observe the correlation effect via impurity hybridization function for $Ce_{0.24}La_{0.76}Ge$, we performed the calculations on two structures. The unit cell of first structure (structure 1) is same as the conventional unit cell. In structure 1, out of four Ce-sites the three sites are replaced by the La atoms. In second structure (structure 2), volume of the unit cell is twice of the volume of the conventional unit cell and, there are 8 Ce-sites, 6 of Ce-sites are replaced by La atoms. Structure 1 and structure 2 give rise to $Ce_{0.25}La_{0.75}Ge$. On these structures, three calculations are performed at temperatures 50, 100 and 300 K. The imaginary part of the impurity hybridization function ($Im\Delta$ ($i\omega$)) corresponding to these two structures are found to be different at the respective temperatures. We have estimated the difference between imaginary parts of the hybridization functions for these two structures (i.e. $\delta$ $Im\Delta$ ($i\omega$) = $Im\Delta$ ($i\omega$) $_{structure\ 2}$ - $Im\Delta$ ($i\omega$) $_{structure\ 1}$) and plotted – $\delta$ $Im\Delta$ ($i\omega$) as a function of frequency ($\omega$) at 50, 100 and 300 K for $J = 5/2$ (figure S4 of the supplementary information). The observed –$\delta$ $Im\Delta$ ($i\omega$) may be considered as an evidence of the spin-spin correlation present in the compound. With decreasing temperature, -$\delta$ $Im\Delta$ ($i\omega$) increases. This indicates to an increase in spin-spin correlations, which is as per our experimental observation.

Figure 4 (a) and 4 (b) shows the probability of finding $4f$ electrons in the electronic configuration $N$ at temperature 30 K for CeGe and $Ce_{0.24}La_{0.76}Ge$, respectively. For CeGe, it is observed that the most probable electronic configuration is $4f^1$ ($N = 1$) having a probability value of 0.935. The probabilities for the $4f^0$ ($N = 0$), $4f^2$ ($N = 2$), and $4f^3$ ($N = 3$) are the 0.024, 0.039, and $3.04*10^{-4}$, respectively. This observation suggests that the $4f$ electrons are localized and the mixed valence states are almost absent. For $Ce_{0.24}La_{0.76}Ge$, we find the probabilities 0.793 and 0.19 for $4f^0$ and $4f^1$, respectively. With the introduction of La, the probability of finding electrons in $4f^1$ configuration is reduced and that in $4f^0$ configuration is increased. It indicates that $4f$ electrons favor to stay in a mixed valence states in $Ce_{0.24}La_{0.76}Ge$. In order to see the effect the spin fluctuations, we plot the pie-chart, which shows the probability for the $J_z$ state (inset of figure 4(a) and 4(b)) for CeGe and $Ce_{0.24}La_{0.76}Ge$, respectively. For $|J_z| = 0.5$, 1.5, 2.5 and 3.5, a probability of 33.45%, 26.51%, 39.91% and 0.12% are found, for CeGe, whereas, the respective probabilities are 28.81%, 28.98%, 28.73%, and 13.48% for $Ce_{0.24}La_{0.76}Ge$. This shows a presence of strong spin fluctuations in both compounds. In CeGe, the spin fluctuations refer to the fluctuations of the magnetic moments which are in localized state and form the magnetic ordering through conduction electrons [25]. In the case of $Ce_{0.24}La_{0.76}Ge$, the probability of finding electrons in $4f$ orbitals is around 0.19. In this compound, the magnetic ordering was also not observed [24]. Hence, in this scenario it appears that the $4f$ electrons present in the localized state fluctuates. This fluctuation may give rise to NFL state as observed experimentally. The presence of NFL state in this compound can be verified through studying the behavior of self-energy at around the Fermi level [21, 23].

### 3.2.2 Self-energy

We introduce the Green's function of the DMFT approximation in term of self-energy $\sum(\omega)$ [46]

$$G(\omega) = \frac{1}{\omega - \varepsilon_0 - \sum(\omega)} \qquad \ldots\ldots\ldots (6)$$

$\sum(\omega)$ contains the real part of self-energy $Re\sum(\omega)$ and imaginary part of self-energy $Im\sum(\omega)$. For non-interacting systems the pole is $\omega = \varepsilon_0$. This is modified to $\omega - \varepsilon_0 - Re\sum(\omega)$ by electron-electron interactions. For Fermi liquids, in the vicinity of $\omega = 0$ (Fermi level), the real and imaginary self-energy can be expanded as [46]

$$Re\sum(\omega) \sim -\omega + constant \qquad \ldots\ldots\ldots (7)$$

$$Im\sum(\omega) \sim -\omega^2 \qquad \ldots\ldots\ldots (8)$$

$Re\sum(\omega)$ give rise to a negative slope whereas $Im\sum(\omega)$ have a maximum in the vicinity of Fermi level, which can be derived respectively from the equation (7) and (8). For CeGe, figure 5 (a) and (b) show the $Re$ and $Im\sum(\omega)$ at temperature 30 and 100 K, respectively, for $J = 5/2$ of CeGe. It is noted that $Re\sum(\omega)$ give rise to negative slope and $Im\sum(\omega)$ has the dip at around Fermi level at 30 K. Similar feature has also been reported for α- and γ -Ce [22]. The lifetime of the quasiparticle depends on $Im\sum(0)$. In the PDOS results, as the quasiparticle picture appears in this compound in the vicinity of the Fermi level. In case of Fermi liquids, the lifetime of quasiparticles is infinite at the Fermi level [46]. For CeGe, the value of $Z_k|Im\sum(0)|$ (where $Z_k$ is the renormalization factor, depends on the effective mass ($m^*$) such as $1/Z_k \sim m^*$) is around 1.2 meV, suggesting that it has a finite scattering among quasiparticles at Fermi level. As a consequence, it is supposed that the nature of $4f$ electrons may not resemble as that of the Fermi liquid behavior in this compound. In Ref [25], a temperature-dependent resistivity has been reported experimentally for this compound which shows the metallic behavior deviated from description of the Fermi liquid theory

For Ce$_{0.24}$La$_{0.76}$Ge, figure 5 (c) and 5 (d) represents the $Re\sum(\omega)$ and $Im\sum(\omega)$ at temperature 15, 30 and 100 K, respectively, for $J = 5/2$. At 30 K, it is observed that $Re\sum(\omega)$ exhibit a positive slope at about 0.024 eV where $Im\sum(\omega)$ has a dip. This observed feature shows a shift towards the Fermi level at 15 K. Hence, it is expected that, as the temperature is further decreased, the positive slope in $Re\sum(\omega)$ and the dip in $Im\sum(\omega)$ may move more towards the Fermi level and finally it may arise at the Fermi level at lower temperature. However, in our case, calculations are hard to run below 15 K due to computational limits. This observed behavior is opposite to that reported for the Fermi fluid theory, in both $Re$ and $Im\sum(\omega)$. Also, the quasiparticle peak is not observed in PDOS results. Hence based on the above observations it can be concluded that there is a presence of NFL state in $Ce_{0.24}La_{0.76}Ge$, which is also in accordance to our experimental reports [24]. We would also like to mention that similar theoretical investigations about NFL state have also been carried out in Ref [21].

Further to see the implications of the development of the NFL state in $Ce_{0.24}La_{0.76}Ge$, it is useful to introduce Matsubara self-energy, in which all interaction effects are squeezed. Moreover, Mastubara self-energy does not carry the errors arising due to the use of maximum entropy method for analytical continuation of self-energy from imaginary to real frequency. In the NFL state, quantum fluctuations mediate the fermionic-fermionic interactions, which yield a Matsubara self-energy $Im\sum(\omega) \propto \omega^{\alpha}$ with $\alpha < 1$ [19, 20]. Matsubara self-energy can be obtained from the interacting Green's function [46]

$$\mathcal{G} = \frac{1}{i\omega - \varepsilon_0(k) - \sum(i\omega)} \dots\dots\dots (9)$$

In this equation, the imaginary part of Matsubara self-energy $\sum(i\omega)$ varies as [19, 20]

$$Im\sum(i\omega) = A\ (i\omega)^{\alpha} + \gamma \quad \dots\dots\dots (10)$$

In the Refs [19, 20, and 23] it has been reported that $Im\sum(i\omega)$ should exhibit a linear behavior at low energy for Fermi liquids and intercept $\gamma$ is related to the quasiparticle mass enhancement. In Fig. 6, we have plotted $\omega$ dependent $-Im\sum(i\omega)$ and we observe $\alpha = 0.42$ for $J = 5/2$. This non-linear $\omega$ dependence of the $-Im\sum(i\omega)$ implies to the existence of the NFL state which is in analogy with the above results discussed for this compound.

Finally, based on the self-energy data, the effective mass $m^*$ can be estimated from the equations [46]

$$m^* = m_e(1 - \frac{\delta Re\sum(\omega)}{\delta\omega})|_{\omega=0} \quad \dots\dots\dots (11)$$

where $m_e$ is the mass of the non-interacting band electron. For CeGe, the theoretically calculated effective mass is $\sim 374\ m_e$ for $J = 5/2$ and $\sim 1.54\ m_e$ for $J = 7/2$, at 30 K (see figure S5 of supplementary information where temperature dependent of $m^*$ is plotted). For $J = 5/2$, the calculated effective mass is closer to the experimental value ($\sim 452\ m_e$, which is calculated from Sommerfeld coefficient $\gamma = 433$ mJ/mol-$K^2$ [25]), suggesting that large value of $\gamma$ arise from the $J$ = 5/2 sub-bands. For $Ce_{0.24}La_{0.76}Ge$, the observed values of the $m^*$ for the $J$ = 5/2 is the about 1.06 $m_e$, at 30 K (see inset of figure S5 of supplementary information). For $J$ = 7/2, the values of $m^*$ is similar which is around 1.08 $m_e$, at 30 K. It suggests that effective mass is orbital independent in this compound which might be due to the small contributions of $4f$ electrons as a result of 76% La at Ce-site. However, this theoretical value is close to the obtained experimental value of 3.87 $m_e$, which is derived from $\gamma$ (= 2.43 mJ/mol-K$^2$ taken from Ref [24]).

## 4. Discussion

In this section, we attempt to explain the bi-quadratic exchange coupling through Heisenberg model in the $Ce_{0.24}La_{0.76}Ge$. But before that, we first try to understand the model for magnetic state of the CeGe whose structure is similar to $Ce_{0.24}La_{0.76}Ge$ compound. CeGe crystallize in the orthorhombic structure with space group *Pnma*, which contains four Ce-sites per unit cell. In this structure, the Ge atom is located in a triangular prism of the Ce atoms (see figure S6 of supplementary information). The structure is similar to FeB-type structure [56]. In each Ce atom, $4f$ shell with one electron ($Ce^{3+}$) have $S$ = 1/2, $L$ = 3, $J$ = |$L \pm S$| = 5/2, 7/2 (e.g., ↑, 0, 0, 0, 0, 0, 0). At low temperatures, it is expected that $Ce^{3+}$ multiplet $4f^1$ is separated into doublet and a quartet state, instead of three doublets, under the influence of the crystalline electric field (CEF) [25, 47]. The quartet state has been observed in other Ce-based compound, e.g. $CeB_6$, in their high symmetry cubic crystal field [48, 50, and 57]. In the low symmetry structure, e.g. orthorhombic structure, it has been discussed that weak CEF effect can be responsible for the presence of quartet state [47]. Similarly, the weak CEF, as pointed out, seems to be the most probable cause for the presence of quartet state in the case of the CeGe. As a result in this compound, Ce-$4f$ electrons undergo two ordering phenomena, one driven by dipole-dipole interactions and the other by interactions between multipolar moments [25]. To understand the possible mechanism we write the Heisenberg model, in ordered state, for this compound [58-60]

$$H = \sum_{i,\delta_n} \{J_n S_i . S_j + K_n (S_i . S_j)^2\} \quad \ldots\ldots (12)$$

where $j = i + \delta_n$ ($\delta_n$ connects site $i$ and its $n^{th}$ nearest neighbor sites). $J_n$ and $K_n$ are the bi-linear and bi-quadratic couplings between the $n^{th}$ nearest neighbor spins, respectively. These couplings undergo the sharp magnetic transition, as seen in temperature response of magnetic susceptibility and heat capacity measurements [25]. This suggests that $n$ should be greater than 1. In experimental study, the results are reported for polycrystalline sample; therefore we limit our consideration to isotropic bi-linear and bi-quadratic exchange terms. The bi-quadratic exchange term can be expressed as (which has proposed for local effective magnetic moment $S \geq 1$) [58-60]

$$(S_i . S_j)^2 = 1/2 Q_i . Q_j - 1/2 S_i . S_j + 1/3 S_i^2 S_j^2 \ \dots (13)$$

where $Q_i$ and $Q_j$ are the quadrupolar moment at site $i$ and $j$, respectively. By inserting equation (13) into (12), the equation (12) can be rearranged as

$$H = \sum_{i,\delta_n} \{(J_n - K_n/2) S_i . S_j + K_n \ (1/2 Q_i . Q_j + \ 1/3 S_i^2 S_j^2) \} \ \dots (14)$$

Introduction of La at Ce-site of CeGe suppresses magnetic ordering and develops NFL state in $Ce_{0.24}La_{0.76}Ge$ compound [24]. NFL state is successfully described by DFT + DMFT method. The results of DFT + DMFT calculations are similar to the physics of analytically solvable Sachdev-Ye model. Also according to this model, in NFL state Matsubara self-energy varies as $\sqrt{\omega}$ [61, 62]. It is implied that such an NFL state is possible due to strong local moment fluctuations from unquenched magnetic moments. In this work our theoretical studies on $Ce_{0.24}La_{0.76}Ge$ show the presence of the NFL state associated with local moment fluctuations. This is because La-substitution can perturb the crystal field environment which in turn modifies the CEF states (splitting energy and level scheme). Therefore, in $Ce_{0.24}La_{0.76}Ge$, it is expected that the quartet state is destroyed or degeneracy is lifted. The NFL state may originate from the fluctuating nearest neighbor moment ($n = 1$), where $K_n$ and $J_n$ are supposed to be weak. Interestingly, non-linear susceptibility scaling suggests a transition from NFL state to a partial order parameter (associated with bi-quadratic exchange coupling) in the presence of external magnetic fields. To explain such behavior, we take the idea from the reports of Slonczewski *et al.*, [16] and Vlasko-Vlasov *et al.*,[17] which suggests that magnetic fluctuations can be responsible for the development of field-induced bi-quadratic exchange coupling. Similarly, in our case, we can state that local moments interact spatially through conduction bands in the presence of magnetic fields resulting in magnetic fluctuations. From equation (14), we can understand that $K_n$ is expected to play an important role in the presence of magnetic fields. This leads to the development of the bi-quadratic exchange coupling in $Ce_{0.24}La_{0.76}Ge$. As a result, a partial order parameter grows around $T*\pm 1$ (above $2 \ (10^4)$ Oe) (See figure S3 of the supplementary information). Hence, in this compound, it is expected that the bi-quadratic exchange coupling arises due to spatial magnetic fluctuations, origin of which is completely different from the reported higher order magnetization in CeGe.

In addition, from the temperature dependent heat capacity at different fields (figure 3 of supplementary information); the possible phases present in $Ce_{0.24}La_{0.76}Ge$ are extracted, as shown in *H-T* phase diagram (figure 7). This compound has three phases; phase I: NFL state, phase II: paramagnetic (PM) phase and phase III) partial ordering (PO) associated with bi-quadratic exchange coupling. Generally, the Fermi liquid state is formed in the presence of external parameters (like magnetic fields, pressure etc) due to the recovery of the quasiparticle picture. But in $Ce_{0.24}La_{0.76}Ge$, an opposite effect is noted which indicate a relationship between NFL state, magnetic fluctuations, and bi-quadratic exchange coupling. The observed opposite effect is possibly due to the developing spatial magnetic fluctuations with magnetic fields, which develop the bi-quadratic exchange coupling.

## 5. Summary

To summarize, our combined theoretical and experimental study for $Ce_{0.24}La_{0.76}Ge$ compound indicate that, in zero field, there is a presence of local moment fluctuations in the NFL state. In the presence of magnetic fields, local moments interact spatially through conduction electrons and results in the development of bi-quadratic exchange coupling. This exchange coupling is responsible for the appearance of a partial order parameter in heat capacity. The bi-quadratic exchange coupling is derived from spatial magnetic quantum fluctuations. Our work highlights to the fact that the local moments in the NFL state may participate in the development of the high-order exchange coupling rather than, bi-linear exchange couplings or Fermi liquid behavior. This implies an interesting ground state in $Ce_{0.24}La_{0.76}Ge$ which is not in accordance with the proposed Doniach model for other Ce-based compounds. Hence, this work might lead to further microscopic experimental studies to investigate the connectivity between the NFL state and bi-quadratic exchange coupling.

**Acknowledgements**

The authors acknowledge IIT Mandi for experimental and computational facilities and financial support.

**Figures:**

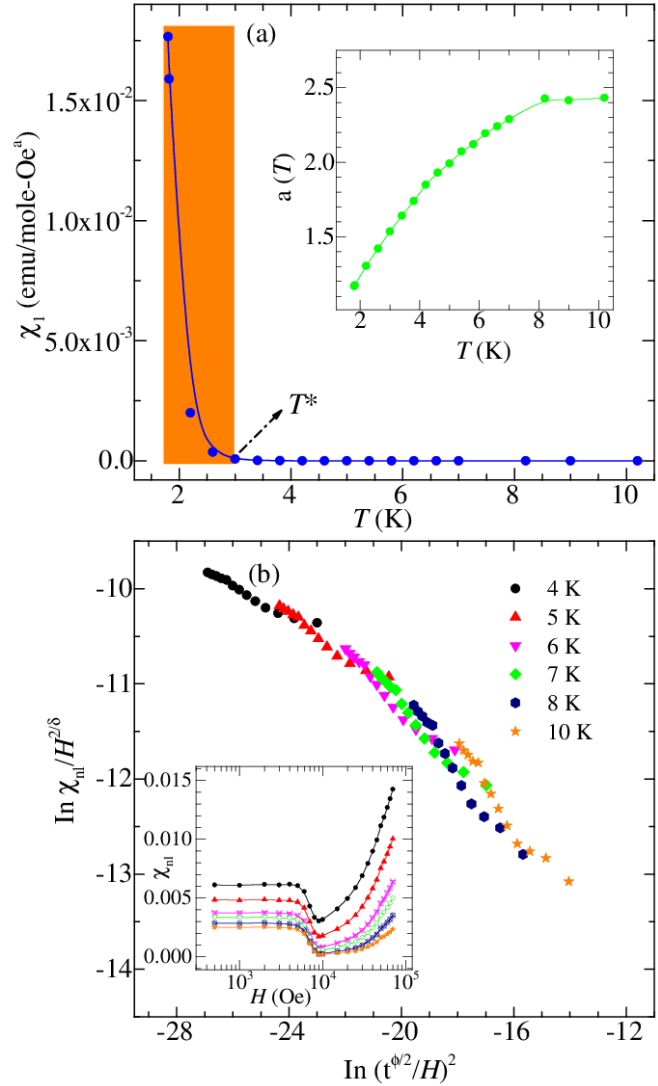

**Figure 1:** (a) Temperature dependent $\chi_1$ for $Ce_{0.24}La_{0.76}Ge$. Orange shaded area represents the higher order susceptibility regime, below temperature $T^*$. Inset: Exponent a($T$) plotted as a function of temperature. (b) Scaling plots of $\chi_{nl}$, above $2*(10^4)$ Oe for $Ce_{0.24}La_{0.76}Ge$. Inset: $H$ dependent $\chi_{nl}$ at different temperatures.

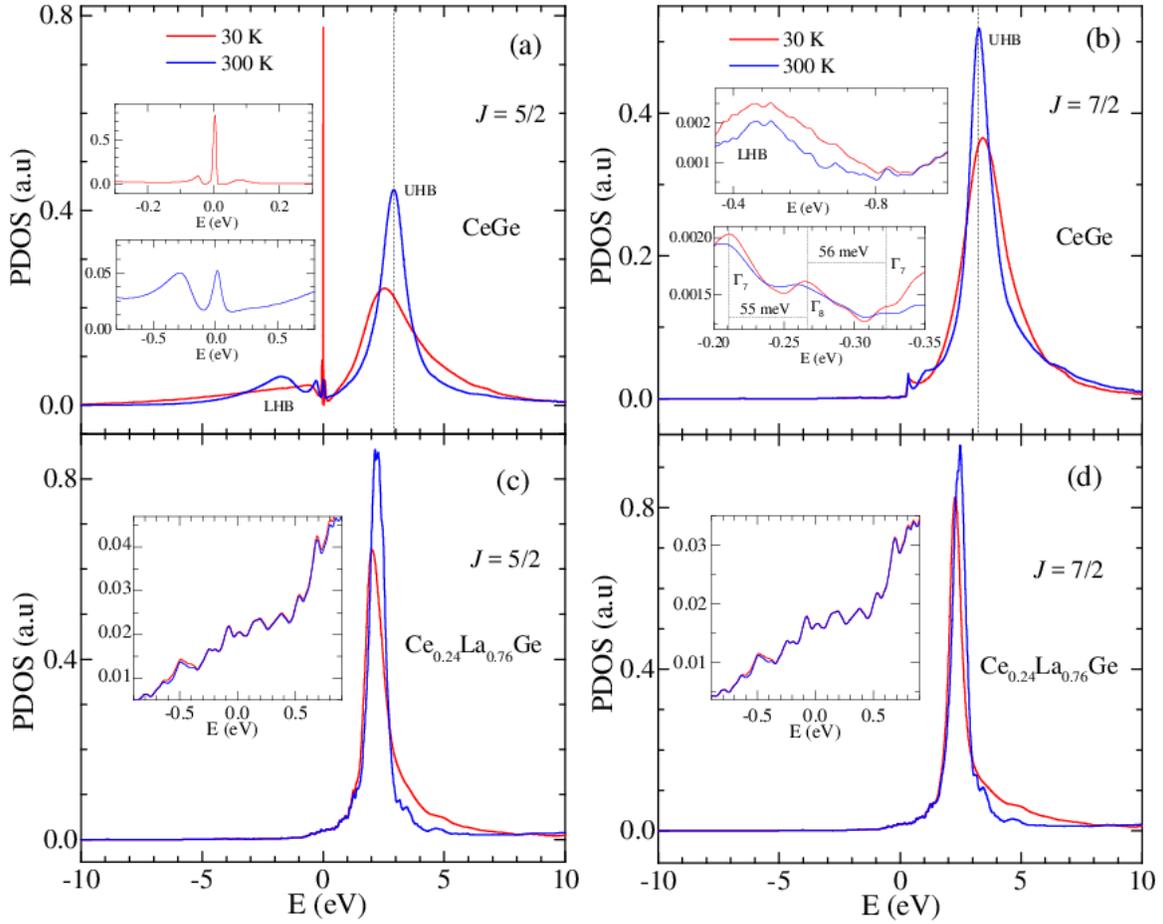

**Figure 2:** Partial density of states (PDOS) of Ce-4*f* at 30 and 300 K: (a) and (b) for CeGe; (c) and (d) for Ce$_{0.24}$La$_{0.76}$Ge. Upper and lower inset of figure (a) shows the PDOS around Fermi level at 30 and 300 K, respectively. Upper and lower inset of figure (b) shows the PDOS at low negative energy at both 30 and 300 K. Inset of the figures (c) and (d) shows the PDOS around Fermi level. Zero energy corresponds to the Fermi level. LHB is the lower Hubbard band and UHB is the upper Hubbard band.

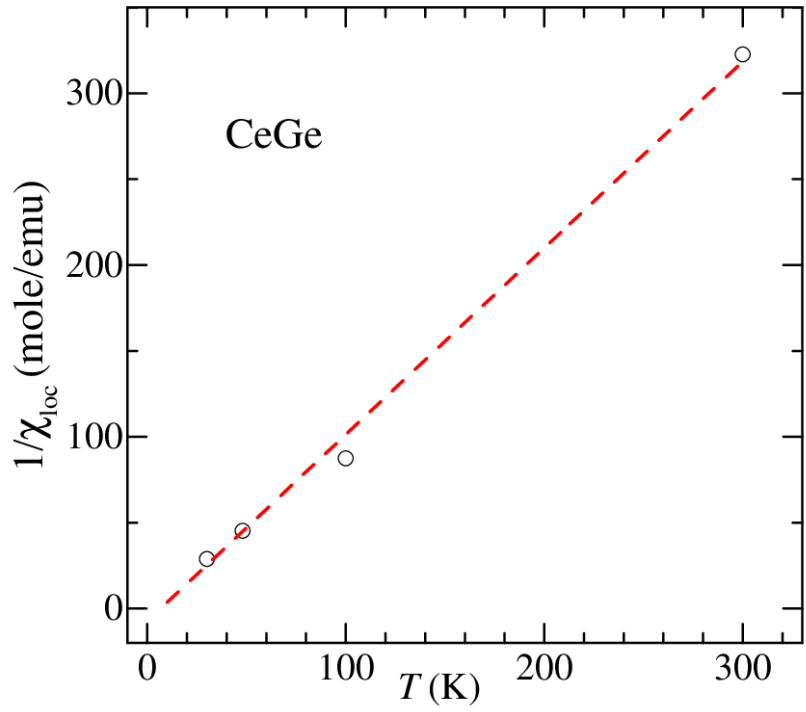

**Figure 3:** Temperature dependent inverse local spin susceptibility ($1/\chi_{loc}$) for CeGe. Red line through the data points indicates to the linear fitting.

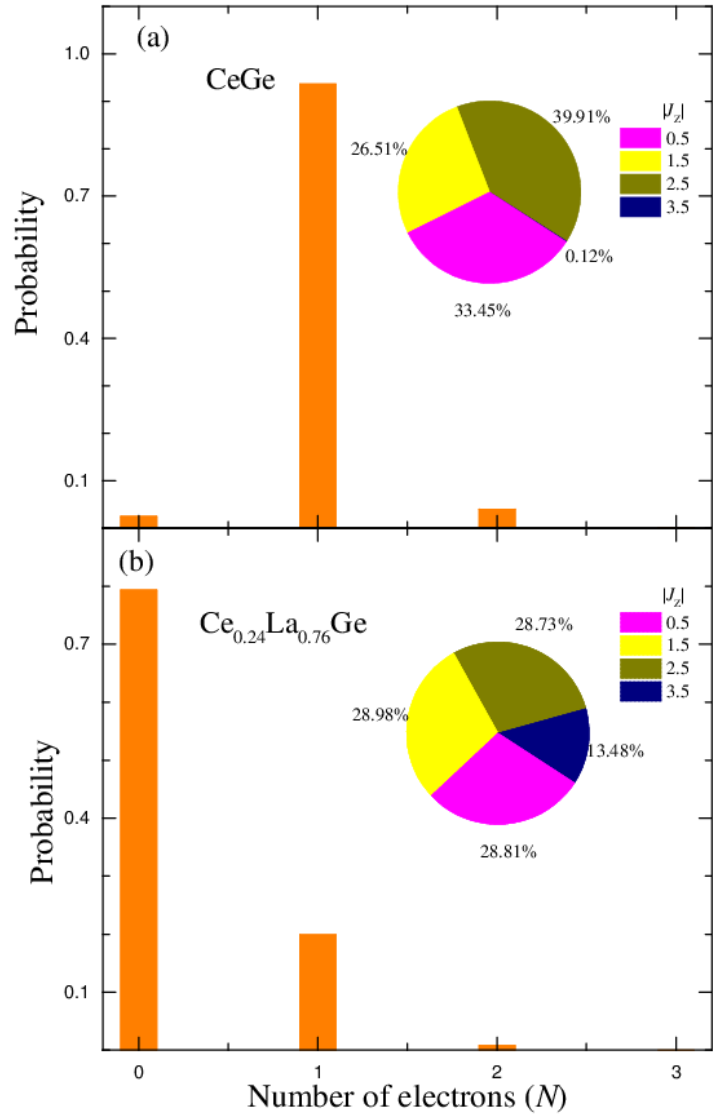

**Figure 4**: Probability of *N* electrons in 4*f*-orbitals at 30 K for a) CeGe and b) Ce$_{0.24}$La$_{0.76}$Ge. Inset of (a) and (b) shows the pie-chart, indicating probability of finding the electron in *J*$_z$ states at 30 K for CeGe and Ce$_{0.24}$La$_{0.76}$Ge, respectively.

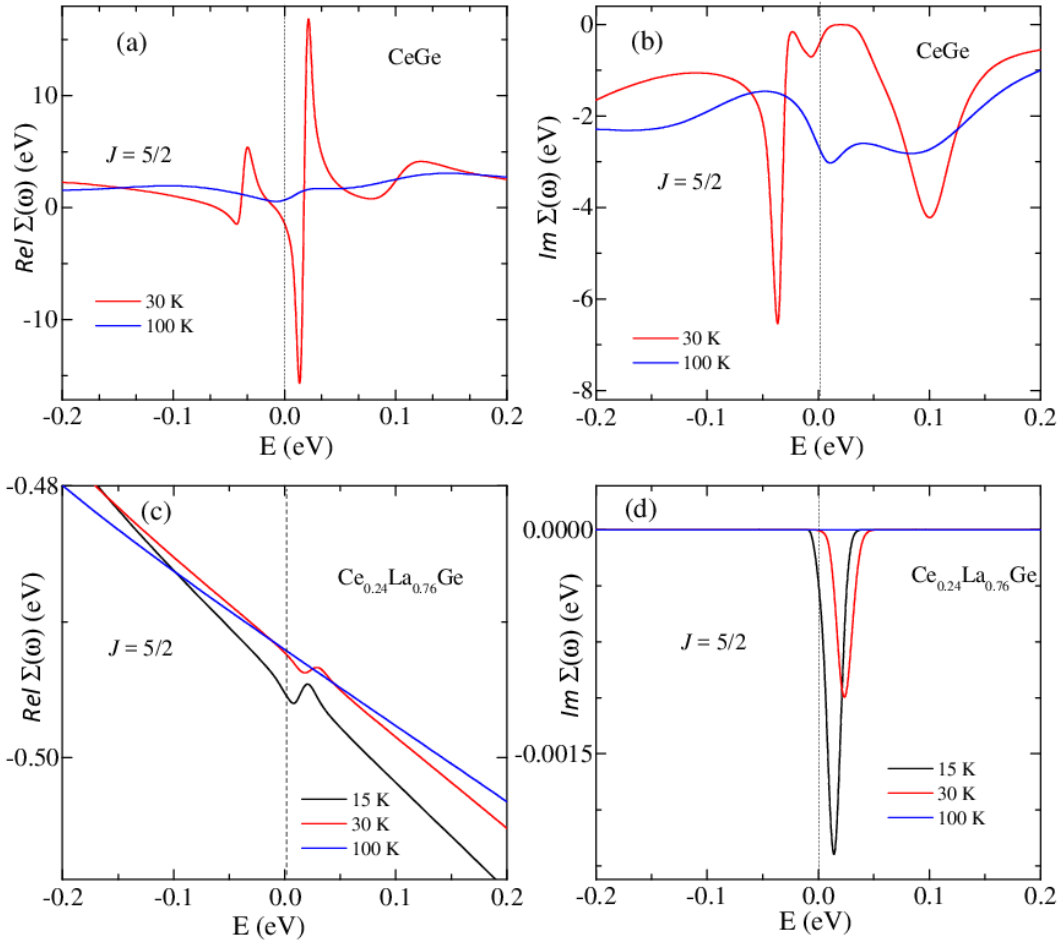

**Figure 5:** (a) and (b) Real and imaginary part of self-energy for $J = 5/2$ at 30 and 100 K for CeGe. (c) and (d) Real and imaginary part of self-energy at 15, 30 and 100 K for $Ce_{0.24}La_{0.76}Ge$. Zero energy corresponds to the Fermi level.

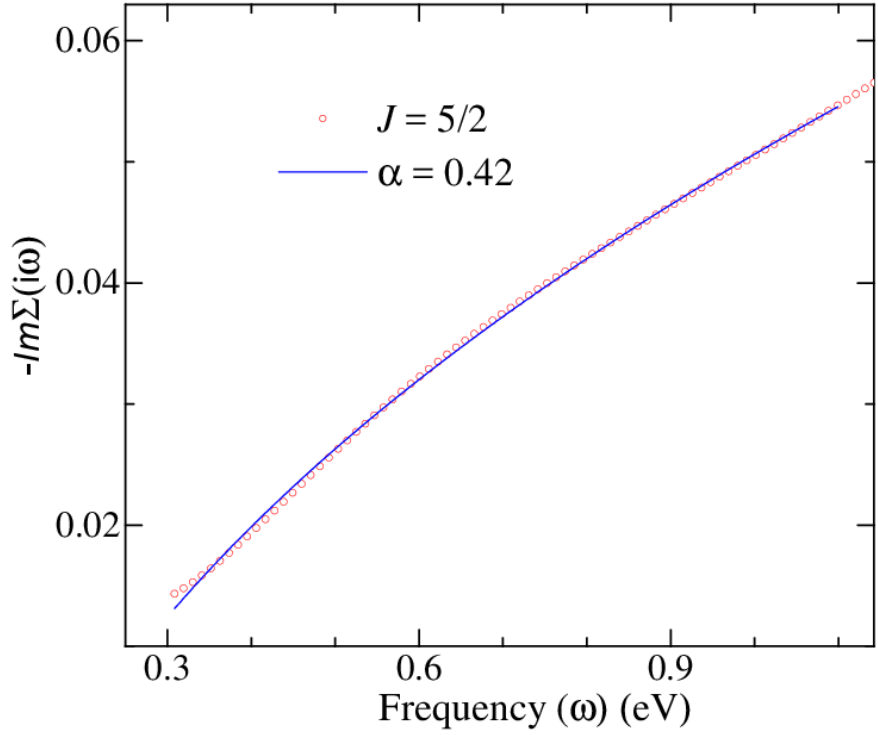

**Figure 6:** Imaginary part of the Matsubara self-energy function $\sum(i\omega)$ for $J = 5/2$ at 20 K for $Ce_{0.24}La_{0.76}Ge$. The solid line indicates the fitting using equation (10).

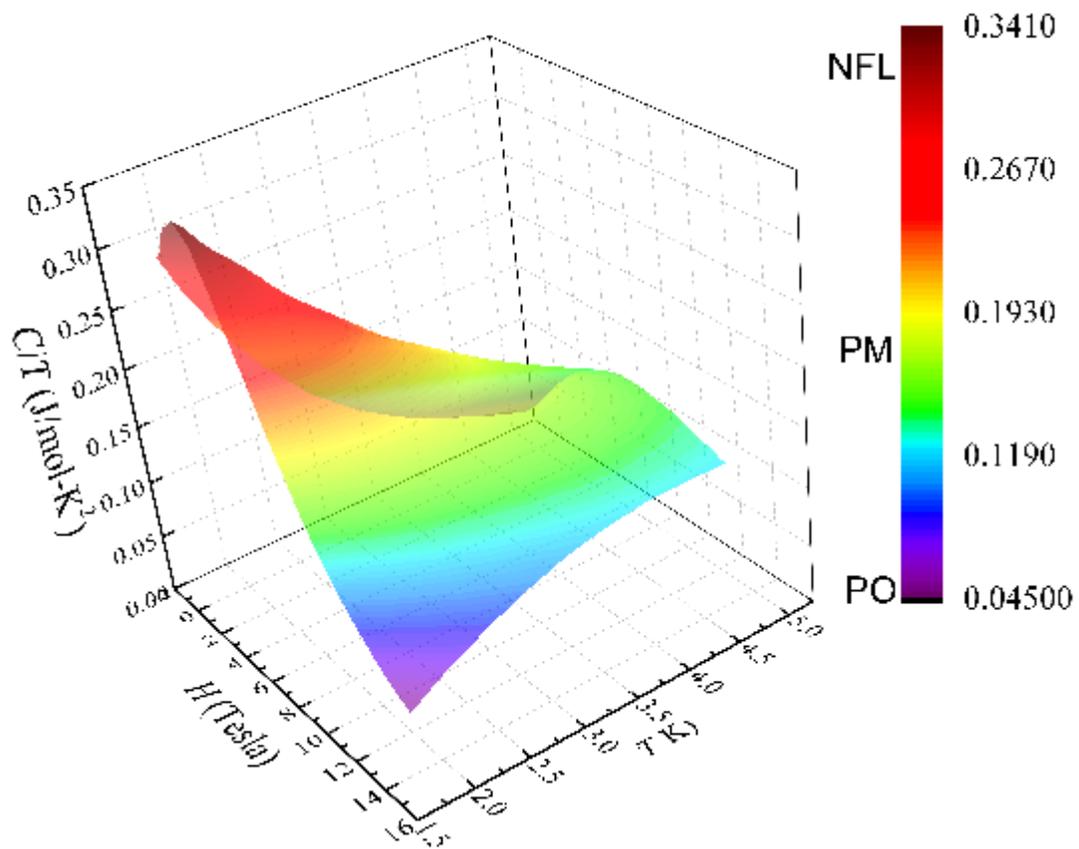

**Figure 7.** *H-T* phase diagram extracted from the temperature dependent heat capacity at different fields for Ce$_{0.24}$La$_{0.76}$Ge. This diagram indicates to three phases 1) NFL state with short-range correlations, 2) paramagnetic PM phase and 3) partial ordering (PO) associated with the bi-quadratic exchange couplings.

# Supplementary Information

**Coexistance of non-Fermi liquid behavior and bi-quadratic exchange coupling in La-substituted CeGe: Non-linear susceptibility and DFT + DMFT study**

Karan Singh[1], Antik Sihi[1], Sudhir K. Pandey[2] and K. Mukherjee[1]

[1]School of Basic Sciences, Indian Institute of Technology Mandi, Mandi 175005, Himachal Pradesh, India

[2]School of Engineering, Indian Institute of Technology Mandi, Mandi 175005, Himachal Pradesh, India

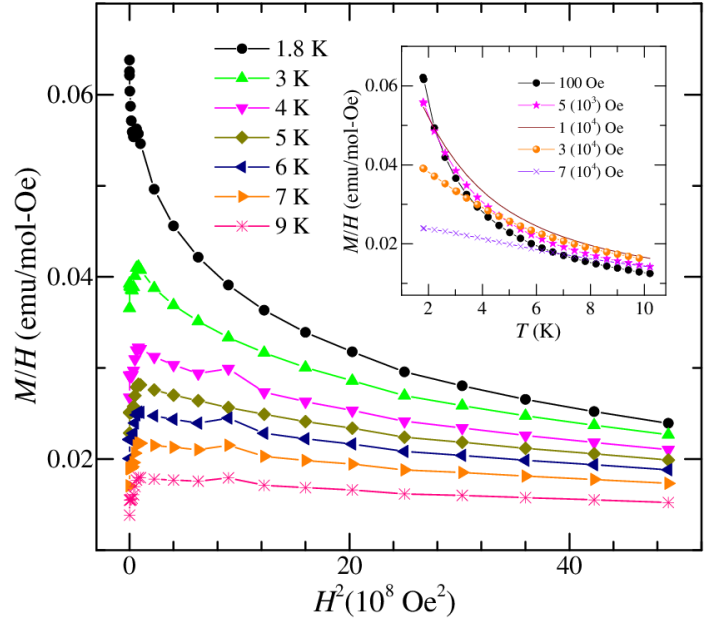

**Figure S1:** $H^2$ dependent $M/H$ at different temperatures for Ce$_{0.24}$La$_{0.76}$Ge. Inset: Temperature dependent $M/H$ at different fields.

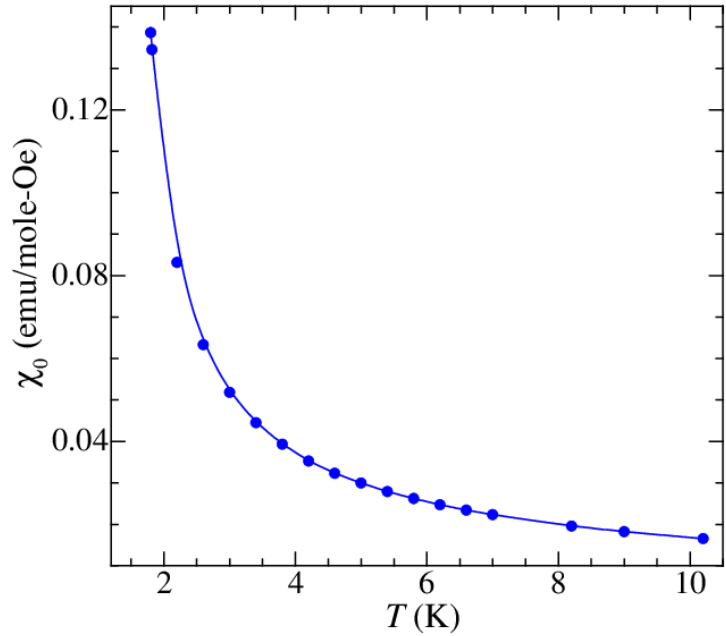

**Figure S2:** Temperature dependent $\chi_0$ for Ce$_{0.24}$La$_{0.76}$Ge.

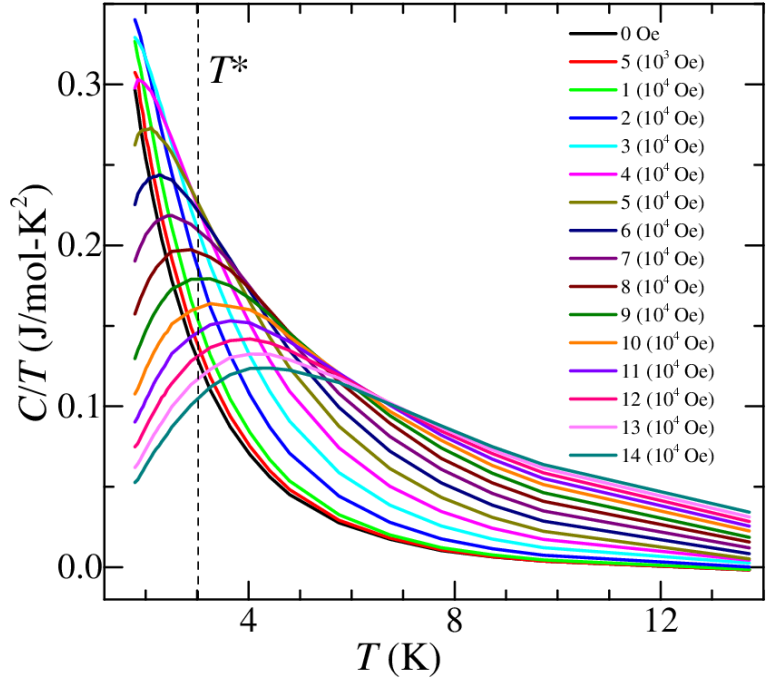

**Figure S3:** Temperature dependent heat capacity at different magnetic fields for $Ce_{0.24}La_{0.76}Ge$ compound. The curves for magnetic fields 0, 0.5, 1, 2, 3, 5, 8, 10, and 14 ($10^4$) Oe are taken from the Ref [Euro Phys. Letts. **126**, (2019) 57005]. $T^*$ is the interaction energy temperature observed from the non-linear susceptibility measurements.

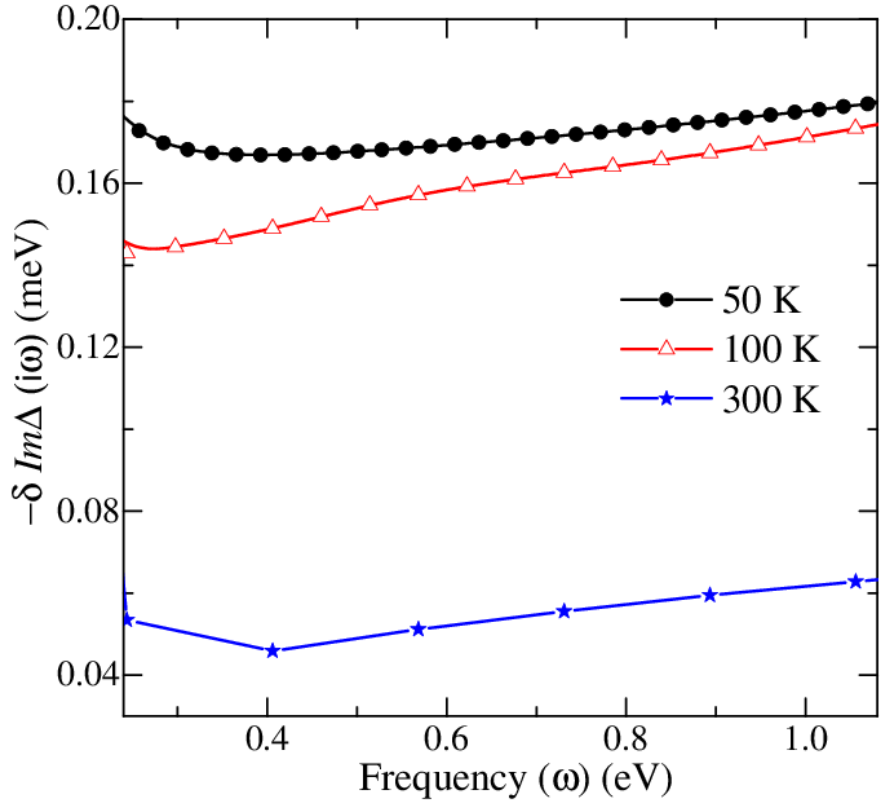

**Figure S4:** - δ $Im\Delta$ (iω) as a function of matsubara frequency (ω) at 50, 100 and 300 K for $J$ = 5/2 for the Ce$_{0.25}$La$_{0.75}$Ge (for details see text).

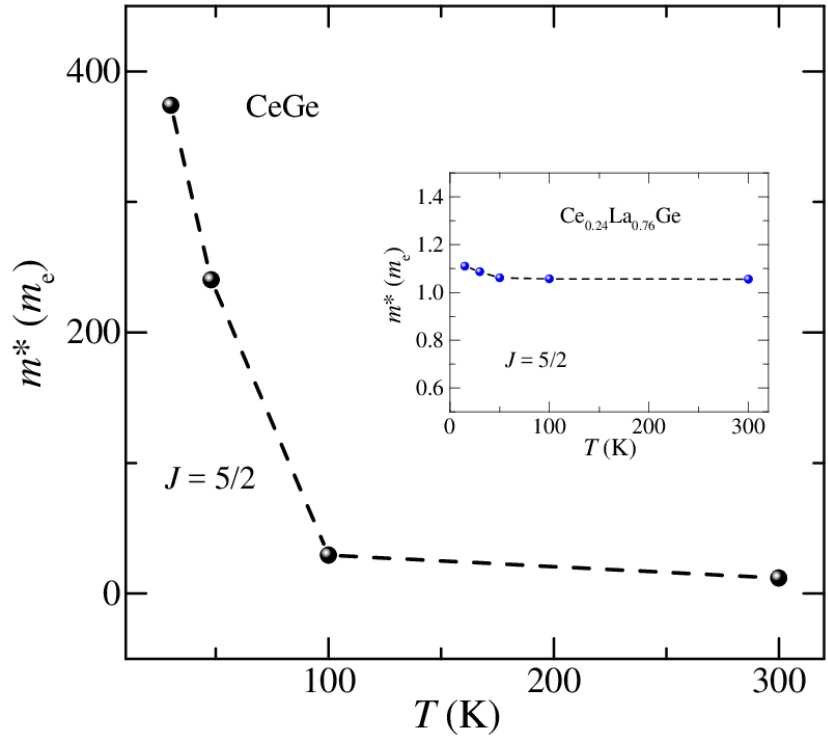

**Figure S5:** Temperature dependence effective mass ($m^*$) for $J = 5/2$ of CeGe. Inset: Similar plot for $Ce_{0.24}La_{0.76}Ge$.

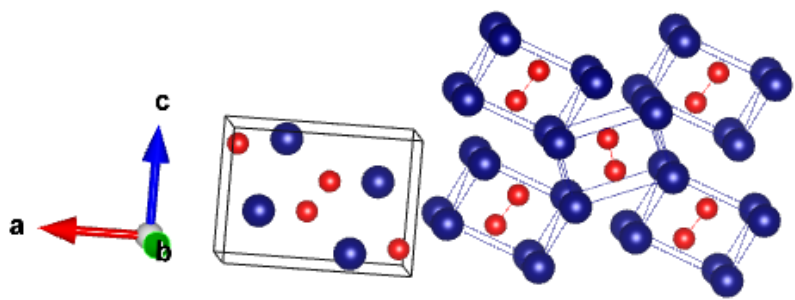

**Figure S6:** The crystal structure of CeGe. Blue and red atom represents Ce and Ge repectively.